\documentclass{interact}

\usepackage{epstopdf}
\usepackage[caption=false]{subfig}
\usepackage{apacite}
\usepackage[colorlinks=true,allcolors=blue]{hyperref}%

\usepackage{soul}

\usepackage[longnamesfirst,sort]{natbib}
\renewcommand\bibfont{\fontsize{10}{12}\selectfont}

\theoremstyle{plain}

\theoremstyle{definition}

\theoremstyle{remark}

\begin{document}


\title{The algorithmic nature of song-sequencing: statistical regularities in music albums}

\author{\name{Pedro Neto\textsuperscript{a}\thanks{CONTACT Pedro Neto. Email: pdealcan@student.jyu.fi}, Martin Hartmann\textsuperscript{a}, Geoff Luck\textsuperscript{a} and Petri Toiviainen\textsuperscript{a}} \affil{\textsuperscript{a}Center of Excellence in Music, Mind, Body and Brain, Department of Music, Arts and Culture Studies, University of Jyväskylä}}

\maketitle

\begin{abstract}

Based on a review of anecdotal beliefs, we explored patterns of track-sequencing within professional music albums. We found that songs with high levels of valence, energy and loudness are more likely to be positioned at the beginning of each album. We also found that transitions between consecutive tracks tend to alternate between increases and decreases of valence and energy. These findings were used to build a system which automates the process of album-sequencing. Our results and hypothesis have both practical and theoretical applications. Practically, sequencing regularities can be used to inform playlist generation systems. Theoretically, we show weak to moderate support for the idea that music is perceived in both global and local contexts.

\end{abstract}

\section{Introduction}\label{sec:introduction}

\begin{figure}[h!]
 \center
 \includegraphics[width=0.9\columnwidth]{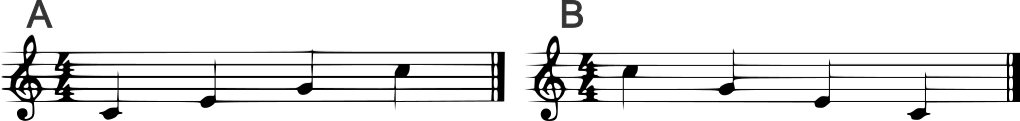}
 \caption{Melody A is different from melody B, even though they are composed of the exact same notes. This illustrates how sequential factors might determine perceptually relevant musical features, such as contour and interval.}
 \label{fig:fig1}
\end{figure}

Music is often described as a time-dependent phenomenon. Contour, for instance, is defined as the pattern of increases and decreases in frequency between consecutive notes. The same is true for rhythms and chord progressions, which are generally defined as patterns of musical events that happen one after the other. For the sake of illustration, consider melodies A and B (\hyperref[fig:fig1]{Figure 1}), which might be readily perceived as different, even though their constituent notes are the exact same.

The recognition of music's sequential nature is responsible for many important findings in the field of cognitive science, as human perception seems to be fairly related to the way that acoustic events happen in a given timeline. The famous probe-tone paradigm showed, for instance, that a single pitch can be perceived as radically different, depending on the notes that were heard before it \citep[- for a review]{krumhansl2010theory}. These results are reliable for many different research paradigms, which generally show that musical context can have a drastic impact on listener behavioral and physiological responses to music \citep{bharucha1987priming, vuvan2011probing, neto2021not, koelsch2007untangling}.

Although fairly robust, these studies have mainly focused on short-term measurements of contextual listening. For instance, a common priming paradigm will assess the influence of a primer on an immediately subsequent stimulus \citep{vuvan2011probing}, or on a stimulus that comes 1 second after the primer \citep{neto2021not}. Still, if we extend the idea of time-dependency to larger musical units, it might be tempting to hypothesize that the experience of listening to a classic sonata cannot be reduced to the experience of simply listening to its constituent movements (although, see section 2 for a discussion).


In fact, music theorists and composers have suggested that a ``motive is heard as part of a theme, a theme as part of a theme-group, and a section as part of a piece'' \citep[p.13]{lerdahl1996generative}. Similarly, André Hodeir (as cited in Lalitte and Bigand, \citeyear{lalitte2006music}, p. 811) states that a ``musical phrase, no matter how beautiful it is, reaches its expressive summit only when it is in perfect harmony with preceding and following phrases''.

The view that music is experienced as a coherent sequence of acoustic events is not an idiosyncrasy of music theorists and classical composers. In contemporary popular music, artists often suggest that there are optimal ways of sequencing tracks in an album, in a party, in a playlist or in a concert. This idea is mainly based, however, on the anecdotal belief that the musical experience can be influenced by the order in which songs are presented to the listener \citep{garvey2013art, nills2015art, ruoff2018how, ruth2019how}. In an overview, artists propose three main concepts of track-sequencing, which can be described as 1) adjacent transitioning, 2) absolute positioning, and 3) overall trajectory. We explore these concepts in the next few paragraphs.

\textbf{Adjacent transitioning} relates to the contrasts and similarities between consecutive songs. It has been suggested, for instance, that pairs of tracks should not be very similar in terms of key and tempo \cite{nills2015art, ruoff2018how, ruth2019how}. According to \cite{nills2015art}, ``I would not choose to put two slow songs next to each other [...] that could get a bit boring''. In an interview with \cite{garvey2013art}, David Brewis goes in the opposite direction, suggesting that ``if you go straight from a quite fast song [...] to just a little bit of a slower song, it can make the slower song seem like it's dragging. We have to avoid that''. In opposition, Guy Harvey adds that ``if you have two songs in the same key, but there is a dramatic tempo increase, that can work'' \citep{garvey2013art}.

\textbf{Absolute positioning} concerns the specific segments (i.e., beginnings, middles and ends) that tracks should assume in an album, depending on their musical characteristics. A common belief is that the first songs of an album are the most relevant for the overall experience of the listener \citep{hahn2018album, wensem2016how, ruoff2018how, sawyer2021how}. This view is sometimes based on the idea that attention spans are short, and that ``if you don’t catch people right off the bat, they might not hear the hits at the end'' \citep{sawyer2021how}. In the same interview to \cite{garvey2013art}, producers Peter Hammill and Ashley Abram agree that there is a traditional view of sticking ``all three hits on the front'' of the album \citep{garvey2013art}. In fact, \cite{friedman2021how} investigated 694 albums from 7 different genres, and found that the first tracks are usually the faster ones.\footnote{This study was published in the author's blog, and it was not scrutinized by peer-review processes.}

\textbf{Overall trajectory}, finally, is the concept that track-sequences should follow some kind of well-established rationale throughout the whole extension of the album. For instance, \cite{wensem2016how} suggests that ``arranging songs by key from lowest to highest'' creates a ``positive feeling'', and that a gradual increase in tempo through an album might evoke a ``rising'' sensation. \cite{hahn2018album} adds that, if sequences of tracks ``build and release tension over the whole release [...] tracks will hit harder individually, and the overall effect [of the album] will be enhanced''.

Although very popular, these ideas have not yet received much attention from the scientific community. We know that musicians and music producers believe in the relevance of album-sequencing. We do not know, however, the extent to which these beliefs actually determine the way that tracks are organized in an album. Here we explore how sequences of tracks are organized within Music Albums (MAs) and, based on the aforementioned anecdotal evidence, we ask the following questions:

\begin{enumerate}
    \item Are there retrievable regularities in the way that tracks are sequenced within MAs?
    \item Are there specific characteristics of tracks in different segments of an album?
    \item Is there a general trend between the first and the last songs of MAs?
    \item Can we use the concepts of adjacent transitioning, absolute positioning and overall-trajectory to automate the process of track-sequencing within MAs?
\end{enumerate}

{\setlength{\parindent}{0cm} These questions are proposed in the context of a broader investigation, where we attempt to elucidate more fundamental issues, such as:} 

\begin{enumerate}
    \setcounter{enumi}{4}
    \item Are there perceptual consequences to different patterns of song sequencing?
    \item To what extent is it true that what we hear now can influence what we hear in the future?
    \item Can we optimize some dimension of the musical experience (e.g., enjoyability or attention) by manipulating the order in which tracks are presented to the listener?
\end{enumerate}

Before we get to the specificities of our investigation, we review two sets of studies focusing on 1) global versus local music perception, and 2) automatic playlist generation. We interpret these studies in light of the questions that were raised here, and we suggest that a deeper understanding of how musicians choose to organize tracks in an album can be beneficial for both of these lines of research.

\section{From sequences of movements to sequences of tracks}\label{sec:prev}

The idea that music is globally perceived as a coherent sequence of acoustic events is a common one amongst composers and musicologists (e.g., \citeauthor{schoenberg1967fundamentals}, \citeyear{schoenberg1967fundamentals}, pp. 1--2; \citeauthor{lerdahl1996generative}, \citeyear{lerdahl1996generative}, 1--5). In contrast, the concatenationist hypothesis suggests that most of our musical experience can be reduced to the perception of local transitions between consecutive musical elements \citep{levinson2006concatenationism, levinson2018music}. Essentially, these opposing views are disputing the extent to which human beings can perceive sequential structures in music. Whereas proponents of a global view believe that individuals appreciate the relationship between movements of a symphony that are minutes or even an hour apart, concatenationists believe that this perceptual process takes place within very short time-windows, which last for only a few seconds.

In contrast with the global-perception hypothesis, the concatenationist view receives significant support from the psychological literature \citep{tillmann2004relative, konevcni1984elusive, gotlieb1985effects, karno1992effects, eitan2008growing, rolison2011role}. \cite{konevcni1984elusive}, for instance, found that randomly scrambling the movements of Beethoven's compositions affected neither the pleasantness nor the emotional impact of these pieces. These findings are consistent for different methods, composers and music genres, \citep{karno1992effects, gotlieb1985effects, cook1987perception, eitan2008growing, rolison2011role}.

It is interesting to note, here, that global and local accounts of music perception can also be scrutinized in the context of albums and playlists. Just like theorists suggest that the overall structure of a classical piece is perceptually relevant, contemporary producers and musicians believe that track-sequencing can ``make it or break it'' for an album \citep{garvey2013art}. Conceptually, there is no reason to believe that global/local relationships studied by music psychologists do not translate to sequences of tracks within an album. If, as suggestted by the studies cited in the previous paragraph, the ordering of movements within a classical composition does not affect its perceived pleasantness or emotional impact, then the ordering of tracks within an album or playlist may also be immaterial.

Still, most studies that falsify the globalist view are fairly idiosyncratic. As highlighted by \cite{bigand2006we}, research has mainly focused on complex variables that are related to harmonic progressions, motivic development, and tonal closure \citep{konevcni1984elusive, gotlieb1985effects, karno1992effects, eitan2008growing}. In this context, it might be fruitful to transpose the global-local discussion to other musical contexts, such as albums and playlists, but also to consider other types of variables, such as loudness, valence, arousal and tempo.

As a last step before we present our methods and results, we review some studies in the field of Automatic Playlist Generation (APG) which, in our opinion, incorporate the essence of the dispute between proponents of global and local views of music perception. As we will see, some of the basic beliefs held by APG researchers are also in line with the artists' beliefs that track-sequencing matters in the context of albums and playlists.


\section{Playlists, algorithms and the craft of track-sequencing}

The main task of an APG system is to optimize some dimension of the musical experience (e.g., attention, enjoyability) by recommending sets of songs to be heard in a sequence. As will be made evident in the next few paragraphs, researchers generally agree that the quality of a given music-recommendation cannot be reduced to the quality of a single song, but rather to the context in which this song is recommended to the listener---a view that is challenged by music psychologists, but cherished amongst theorists, composers, musicians and album producers.

Here we argue that previous APG systems have adopted a mixture of global and local strategies for playlist generation, where both the contrasts between consecutive songs and the overall characteristics of the playlists are taken into account. Finally, empirical data from these studies also suggests that track-sequencing variables might hold some level of perceptual relevance.

\subsection{Local accounts of playlist generation}

The predominant view of next-track recommendation systems is that pairs of songs should be coherent in a given dimension, and various types of similarity measurements have been proposed in the past \citep{bittner2017automatic, flexer2008playlist,  pohle2005generating, pohle2007reinventing, platt2001learning, kamehkhosh2017user, ikeda2017music, jannach2015beyond}. These studies generally assume that intelligent sequencing algorithms would optimize the similarity between pairs of songs, such that the systems' goal is to provide smooth transitions between tracks in position $k$ and $k+1$.

The assumption of similarity is empirically sound, as listeners generally prefer transitions that are explicitly optimized for smoothness, rather than randomly assembled sequences of tracks \citep{ikeda2017music, kamehkhosh2017user}. In addition, professional DJs have been shown to carefully select similar pairs of songs in their sets \citep{kell2013empirical}, although the concept of similarity might vary significantly between studies. 

\cite{sarroff2012modeling} attempted to automatically differentiate true from artificial pairs of songs within MAs. True pairs were defined as sequences of 2 songs that appeared consecutively in an album, whereas artificial pairs were assembled with random permutations of the album. Their model was trained on different subsets of features computed by The Echo Nest (which today is available through Spotify’s web API), and it was able to differentiate true from artificial pairs with a 22.58\% accuracy, against a 20\% baseline. Even though this result was barely above chance levels, it shows some support for the concept of adjacent transitioning, since pairs of songs seem to follow some kind of transition regularity \citep{garvey2013art, nills2015art, ruoff2018how, ruth2019how}. 

Still in agreement with the concept of adjacent transitioning, \cite{baccigalupo2006case} argued that APG systems could replicate co-occurrences of song sequences from manually-generated playlists. For instance, if songs $A$ and $B$ appear together in many different playlists, this pair of songs could be considered as a ``meaningful'' sequence, and could be automatically replicated to improve the quality of automatic recommendations. 

\subsection{Global accounts of playlist generation}

\cite{cliff2000hang} proposed a system where listeners would explicitly determine the global trajectory of the playlist. For instance, a user could set the loudest song at the middle of the playlist, and the quieter one at the end. The system would then rank-order songs increasingly towards the middle, and decreasingly towards the end. This view is similar to the concept of overall trajectory \citep{wensem2016how, hahn2018album}, which states that gradual movements throughout the extension of the album can improve the music-listening experience.

Language models \citep{mayerl2019language, mcfee2011natural, liebman2019right} can incorporate both notions of global and local playlist generation, as they generally compute the conditional probability of a given track as function of the tracks that came before it. \cite{liebman2019right} mapped user satisfaction to different transition states by using reinforcement learning. Again, \cite{liebman2019right} showed that sequence-aware algorithms are better for making next-track recommendations, if compared to systems that disregard timeline factors. Theoretically, language models can function in a global level by allowing higher order models, which will calculate the probability of song $A$ given the $n$ songs that came before it.

If analysed together, these studies reveal that APG researchers generally agree upon the idea that a track is not perceived as an independent musical unit, but rather as a member of a broader context. Also, we see that different approaches to music recommendation can assume global and/or local accounts of music organization, as well as some of the concepts expressed by musicians and producers interested in album-sequencing.

\section{Epistemological notes}

Not unlike previous studies, we make an arbitrary distinction between the concepts of global, local, adjacent and overall musical structures. Conceptually, every segment of an album can be considered as adjacent to another segment, depending on the time-frame that we use as our unit of analysis. Similarly, any time-frame can have a global, or an overall structure spanning from its beginning to its end, no matter how short this time-frame may be.

In psychological studies, for example, sequential relationships are usually considered local if they happen within a window of up to 30 seconds \citep{lalitte2006music}. In our study, however, we conceive a track as the minimal unit of analysis, and we assume that the relationship between consecutive songs is local (or adjacent) even though they occupy windows that are usually longer than 60 seconds.

One could, in fact, argue that a better account of local transitioning should be restricted to smaller time-frames, such as the transition from the end of track $k$ to the beginning of track $k+1$. Whereas we agree with this argument, and are interested in any results which might come from such an analysis, we would still argue that our study does not intend to provide a categorical and absolute account of what is local and what is global, adjacent, or an overall trajectory. Rather, we simply consider the patterns that might exist 1) between consecutive tracks, and 2) throughout sequences of $n$ tracks.

\section{Methods}\label{sec:parsons}

The main goal of the present study is to search for patterns of track sequencing within MAs. Albums were chosen because deciding the order of the songs is practically an unavoidable step in the album-production process. It is fair to assume that, after the long and expensive recording, mixing and mastering of their work, artists will not settle for a random sequence of tracks, and that the final sequence will be a deliberately chosen one.

In addition, those responsible for the release of an album are often the musicians and/or the producers themselves, which guarantees some level of musical expertise throughout the process. Playlists, on the other hand, can be made and modified quickly by anyone, without additional costs, or knowledge barriers. We could not assume, therefore, that playlist sequences are created with the same level of skill and deliberateness as MAs.

Based on the anecdotal evidence that we described here, we were able to analyse MAs with regards to patterns of 1) adjacent transitioning; 2) absolute positioning, and 3) overall trajectory of tracks within MAs. These patterns were investigated by means of 1) building a Markov model to describe transition regularities between adjacent tracks; 2) evaluating the frequency with which different tracks are positioned throughout segments of an album; and 3) evaluating the existence of global trends from the first to the last tracks of each MA. Finally, we created a system which attempts to automate the process of track sequencing within MAs. Below we describe each one of these approaches in detail, as well as the dataset that we used in order to accomplish these goals.

\subsection{Data}

Using Spotify's web API, we queried for artists names from a list of 32 different genres. Each artist was then queried for its complete discography, with a maximum of 50 albums per artist. This method yielded 475342 albums from 26248 artists, which were then filtered in order to remove duplicates, similar versions, and albums with unusually large or small numbers of tracks (i.e., $k < 6$ or $k>16$). The final sample comprised 51010 albums from 8190 artists and 548852 tracks. \hyperref[table:table0]{Table 1} shows the distribution of albums per genre in the final sample.

\begin{table}[]
\tbl{Table 1 - Distribution of albums per genre.}{
\begin{tabular}{rlrr}
  \hline
  & Genre & n & Percentage within the sample \\ 
  \hline
  1 & country & 2874 & 5.63 \\ 
  2 & electronic & 3908 & 7.66 \\ 
  3 & indie & 725 & 1.42 \\ 
  4 & jazz & 7884 & 15.46 \\ 
  5 & latin & 3000 & 5.88 \\ 
  6 & pop & 9504 & 18.63 \\ 
  7 & rap & 1947 & 3.82 \\ 
  8 & reggae & 370 & 0.73 \\ 
  9 & regional & 709 & 1.39 \\ 
  10 & rock & 19329 & 37.89 \\ 
  11 & soundtrack & 760 & 1.49 \\ 
  \hline
\end{tabular}
\label{table:table0}}
\end{table}

Each track within each album was described by a subset of the features provided by Spotify. We chose to work only with valence, energy, loudness, and tempo, discarding features like danceability, speechiness, acousticness and liveness, as these are either conceptually irrelevant to our research question (e.g., liveness), or are hard to interpret and/or replicate outside of Spotify's web API (e.g., danceability). Despite the fact that we do not handle the computation of our musical features, nor have access to the way that they were computed, we chose to work with this dataset because, to our knowledge, it is the only one that 1) provides information about where each track is positioned within an album, and 2) provides complete albums without missing tracks. 


\subsection{Adjacent transitioning}
\subsubsection{Feature representations}

As a way to compute the direction of feature variation between adjacent tracks, we represent song sequences as patterns of ups and downs in a given feature domain (i.e., valence, energy, loudness, and tempo). This approach focuses on the direction of transitions between consecutive notes, in a way that is similar to the Parsons Coding of melodic contour \citep{parsons1975directory}. By making our data discrete, rather than continuous, it  allows us to build a Markov Model to describe transition probabilities between consecutive songs.

We define an album $a$ as a sequence of
tracks $T = (t_1$, $t_2$, \dots, $t_k)$,
where $t_k \in T$ is represented by a vector
of $j$ audio features $F = \{f_1$, $f_2$,
\dots, $f_j\}$. In Parsons coding, we convert
each feature of $F$, to a sequence of states
representing the direction of consecutive
transitions between tracks in positions $k$
and $k+1$. For each album $a \in A$, and
feature $f_j \in F$ where $A$ is the complete
set of albums in our data set, we compute $k'
= k-1$ transitions $t'_{k'j}$ as:

 \[ t'_{k'j} = 
    \begin{cases}
        \text{up}, & \text{if}\ t_{k,j} < t_{k+1,j} \\
        \text{down}, & \text{if}\ t_{k,j} > t_{k+1,j} \\
    \end{cases}
  \]

  After encoding track transitions, each album is represented by $j$ vectors of $k'$ feature transitions, which assume one of $n = 2$ possible states $st_n \in \{up, down\}$. We ignore the possibility of $t_{k,j}$ being equal to $t_{k+1,j}$, as these are highly unlikely for vectors of continuous audio features. Still, one could consider that the feature of a track only goes ``up" or ``down" if the difference between $t_{k,j}$ and $t_{k+1,j}$ surpasses an arbitrarily small threshold. In which case, an additional state of ``same" would be considered. For simplicity, we choose to work with a binary category of ``up" and ``down".

\subsubsection{Transition probabilities}\label{subsec:markov}

Given a sequence of transitions $T'_j =(t'_ {1,j}, t'_{2,j},\dots, t'_{k'j})$ and its corresponding states $st_n$, we are able to compute $P[t'_{k'+1,j} = st \mid t'_{k',j} = st_n]$ which represents the conditional probability that transition $k'$ of feature $f_j$ goes from one state to the other. This allows us to empirically derive $j$ transition matrices $Tm_j$ (Table 2) and to calculate the mean log-likelihood of $T'_j$ as

\begin{equation}
 \label{eqn:eq1}
 \mathcal{L}(T'_j|Tm_j) = \frac{1}{k'} \sum_{i=1}^{_{k'}} \log P[t'_{i+1,j} = st \mid t'_{i,j} = st_n]   
\end{equation}

Consequently, $\mathcal{L}(a|Tm)$ represents the overall log-likelihood of an album $a$ as the average of $\mathcal{L}(T'_j|Tm_j)$ across all features $f_j \in F$. Similarly, $Tm$ is the set of transition matrices for all $f_j \in F$,

\begin{equation}
 \label{eqn:eq2}
 \mathcal{L}(a|Tm) = \frac{1}{j} \sum_{j \in F}\mathcal{L}(T'_j|Tm_j)
\end{equation}

This approach would help us find local---or adjacent---regularities in the way that tracks are sequenced within MAs. Theoretically, transition matrices could be interpreted as the underlying criteria that musicians and music producers use to determine transitions between consecutive tracks. In the same sense, $\mathcal{L}(a|Tm)$ would be a measure of how much an album $a$ corresponds to the patterns found in our set of transition matrices.

\subsection{Absolute positioning}

In order to evaluate the notion of absolute positioning, we calculated the mean values of $f_j \in F$ throughout different segments of an album. Since MAs can have different lengths, we performed $k$-bins discretizations based on tercile values, which transformed track numbers to categorical values corresponding to the \textit{beginning}, \textit{middle}, and \textit{end} of each album. Feature values were normalized within album prior to the discretization process.

\subsection{Overall trajectory}

With regards to global trends of feature variation, we were interested in evaluating whether our albums presented a general trend of increasing or decreasing features. Spearman-Rank correlations were calculated between 1) track numbers and 2) raw feature values of each track, separately for each album.

Positive coefficients indicate that the album is in an \textit{up ramp}, whereas negative coefficients indicate that it is in a \textit{down ramp}. The magnitude of coefficients would indicate, in turn, the strength of the correlations between track number and loudness values. The distribution of up ramps and down ramps throughout the whole dataset would indicate the extent to which album-sequences favor \textit{down} or \textit{up ramps}.

\subsection{Evaluation methods}
\subsubsection{Adjacent transitioning}

Essentially, \hyperref[eqn:eq2]{eq. 2} calculates how likely a given album is to occur if we assume that its sequences of features $T'_j$ were generated by $Tm_j$. We do not know, however, if this assumption is correct, and therefore we need to evaluate the extent to which $Tm_j$ describes sequences of tracks within MAs.

Arguably, if our transition matrices are used to sort tracks within an album, we should be able to find that the sequence of transitions within an original album $a$, is more likely to occur than a random permutation of the same album, which we refer to as $a'$. We test the hypothesis that $\mathcal{L}(a|Tm) > \mathcal{L}(a'|Tm)$. With a paired-samples t-test, we test this idea against the null hypothesis that likelihood values are the same for random and original albums. If the null is rejected, this would indicate that, under the empirically derived transition matrices, the original sequence from $a$ is more likely to occur than its random permutation $a'$.

Of course, our hypothesis should be tested on a set of albums that was not used to empirically derive $Tm$. This would indicate that $Tm$ generalizes to an album that was never ``seen'' by our model. We build the transition matrices (Table 2) on a subset of approximately 80\% of $A$, and evaluate it on the remaining 20\%. In order to avoid spill-over effects of having the same artist in both the training and the testing samples, the split was performed at the level of the artist. Additionally, we perform a cross-validation procedure by training and testing our model on 10 different subsets of $A$. Statistics and graphs displayed in the results section refer to cross-validated results, unless otherwise indicated.

\subsubsection{Absolute positioning}

A three-way Analysis of Variance (ANOVA) is used to investigate the effect of album segment (i.e., begining, middle, end), condition (i.e., randomized versus original album position), and feature (i.e., valence, energy, loudness, and tempo) on normalized feature values.

\subsubsection{Overall trajectory}

Spearman's $\rho$ values between $f_j$ and track number were calculated separately for each album, and also for the randomized albums. In order to evaluate if albums tend to present \textit{up} or \textit{down} ramps, we computed a pairwise t-test between the $\rho$s of an album and its randomized version. Each album contributes, therefore, with $4$ $\rho$ values, one for each feature.


\section{Results}\label{sec:results}
\subsection{Adjacent transitioning}\label{subsec:markov_eval}

\begin{figure}
 \centerline{
 \includegraphics[width=1\columnwidth]{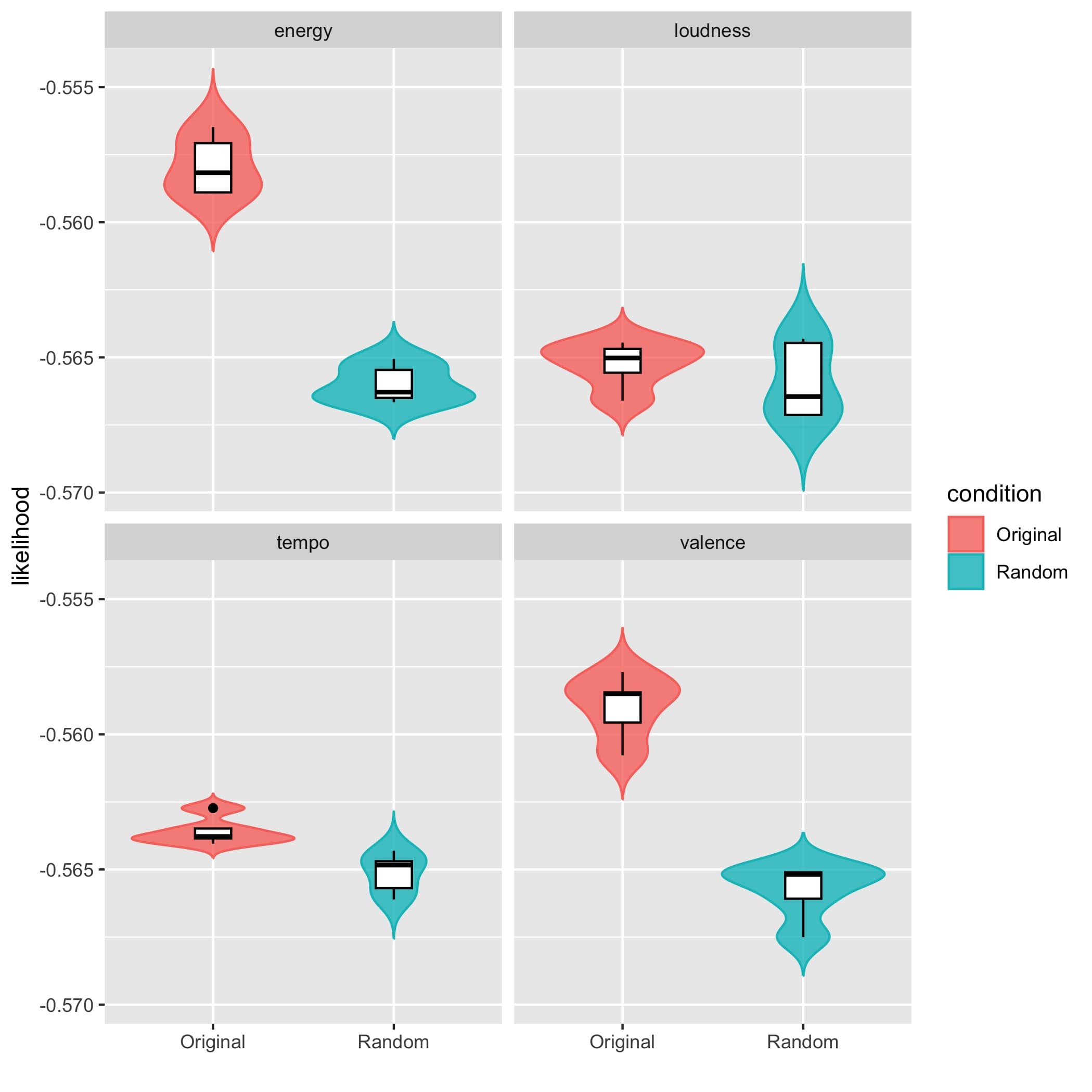}}
\caption{Likelihood of original albums versus random sequences (i.e., $a$ versus $'a$) under the same empirically derived transition matrices (i.e., $Tm$)}
 \label{fig:fig2}
\end{figure}

The paired-samples t-test indicated that, under our empirically derived transition matrices (Table 2), original sequences are more likely to occur than randomized sequences $[t(19)=5.012,\ p<.001,\ d=1.79]$. The difference in likelihood for each feature was evaluated with a set of Tukey corrected t-tests. As shown in \hyperref[fig:fig2]{Figure 2}, the difference in likelihood between random and original sequences was significant for all features, with a less pronounced difference for loudness and tempo ($p<.001$).

\subsection{Absolute positioning}

\hyperref[fig:fig2]{Figure 3} indicates that higher values of valence, energy and loudness are present at the beginning of the each album, but only for the original condition. Differences between album segments are again less pronounced for tempo.

As shown in \hyperref[fig:fig3]{Figure 3}, results of the ANOVA indicated a main effect of album position [$F(2, 4678)=2586.3, p<.001, \eta_{p}^{2}=.00119$], and interactions between condition and album position [$F(2, 4759)=2631.5, p<.001, \eta_{p}^{2}=.00118$], album position and feature [$F(6, 1279)=235.8, p<.001, \eta_{p}^{2}=.000321$], and conditionn, album position and feature [$F(6, 1393)=232.1, p<.001, \eta_{p}^{2}=.00035$].

\begin{table}[]
\tbl{Transition matrices representing the probability of going from the state in a row to the state in a column.}{
\begin{tabular}{llllll}
    \hline
    \hline

    {Valence} & {}  & {}      & {}\vline\vline & {Energy} & {}   \\
    \hline
    {}    & {\textbf{down}}   & {\textbf{up}}   & {}\vline\vline & {\textbf{down}}  & {\textbf{up}}     \\ 
    \textbf{down}  & 0.309     & 0.690   & {}\vline\vline & 0.299    & 0.700     \\
    \textbf{up}    & 0.664     & 0.334   & {}\vline\vline & 0.664    & 0.335     \\
          &          &        & {}\vline\vline &         &          \\
    {Loudness}& {}   &        & {}\vline\vline & {Tempo} &          \\          
    \hline
    {}    & {\textbf{down}}   & {\textbf{up}}   & {}\vline\vline & {\textbf{down}}  & {\textbf{up}}     \\ 
    \textbf{down}  & 0.310      & 0.689  & {}\vline\vline & 0.328    & 0.671      \\
    \textbf{up}    & 0.654      & 0.345  & {}\vline\vline & 0.666    & 0.333      \\ 
    \hline
    \hline
\end{tabular}\label{table:table1}}
\end{table}

\subsection{Overall trajectory}

Down ramps were more frequent than up ramps for all the features, with a less pronounced effect for tempo (\hyperref[fig:fig4]{Figure 4}). Results of the t-test indicated that Spearman $\rho$s are smaller for original albums (M = .073, SD = .34) than for randomized ones (M = .000, SD = .33) [$t(407871) = -67.89, p < .001,\ d=-0.217$].



\section{Automated track sequencing}\label{subsec:sa_eval}

If notions of adjacent transitioning, absolute positioning and overall trajectory are really used by human beings who undertake the task of album-sequencing, we should also be able to use it in order to automate this process. This section describes the rationale and the strategies that we adopted to build such a system.

Assuming that $Tm$ reflects transition regularities, we should be able to use $\mathcal{L}(a|Tm)$ as an objective function to sort adjacent tracks within an album. Consider, again $a'$ as an unordered set of tracks pertaining to an album $a$. Then, we should be able to find a permutation of $a'$ which maximizes its likelihood (\hyperref[eqn:eq1]{eq. 2}). This would arguably approximate $a'$ to $a$, and we could evaluate the quality of $Tm$ as the degree to which $a'$ approximates $a$ in a given dimension.

\begin{table}[]
\tbl{Normalised feature values per feature and album position.}{
\begin{tabular}{rlllr}
 &  &  & Original & Random\\ 
  \hline
 & Album position & feature & mean (SE) & mean (SE)\\ 
 \hline
 & Beginning & valence & 0.067 (0.0021) & 0.000 (0.0021)\\ 
 & Middle & valence & 0.007 (0.0023) & 0.000 (0.0023)\\ 
 & End & valence & -0.080 (0.0023) & 0.001 (0.0022)\\ 
 \hline
 & Beginning & energy & 0.10 (0.0020) & 0.000 (0.0021) \\ 
 & Middle & energy & -0.008 (0.0023) & 0.000 (0.0023)\\ 
 & End & energy & -0.108 (0.0023) & 0.001 (0.0022)\\ 
 \hline
 & Beginning & loudness & 0.118 (0.0020) & 0.000 (0.0021)\\ 
 & Middle & loudness & -0.008 (0.0023) & 0.000 (0.0023)\\ 
 & End & loudness & -0.121 (0.0023) & 0.000 (0.0022)\\ 
 \hline
 & Beginning & tempo & 0.011 (0.0021) & 0.000 (0.0021)\\ 
 & Middle & tempo & 0.003 (0.0023) & 0.003 (0.0023)\\ 
 & End & tempo & -0.015 (0.0022) & -0.002 (0.0022)\\ 
\end{tabular}
}
\end{table}

In order to incorporate the notions of absolute positioning and overall trajectory, we used Spearman's $\rho$ as a penalty term to \hyperref[eqn:eq1]{equation 1}. If the optimal solution based on $Tm_j$ yielded a sequence of tracks in a \textit{down ramp} (i.e., negative correlation between track numbers and feature values) there would be no penalty, with $\rho = 0$. Conversely, if correlations were positive, $\mathcal{L}(T'_j|Tm_j)$ the solution would be penalized by subtracting $\mathcal{L}(T'_j|Tm_j)$ with $\mathcal{L}(T'_j|Tm_j)\rho$. This approach would result in high penalties for strong positive correlations, and no penalty for negative ones (e.g., $\rho = 1$ would result in a likelihood of $0$). We do not represent patterns of absolute positioning explicitly because, by optimizing for negative correlations, we already provide an incentive for our algorithm to position high feature values an the beginning of the album.\footnote{In the discussion of the paper, we expand on the idea that concepts of overall trajectory and absolute positioning revealed to be redundant}

\begin{figure}[h]
 \centerline{
 \includegraphics[width=1\columnwidth]{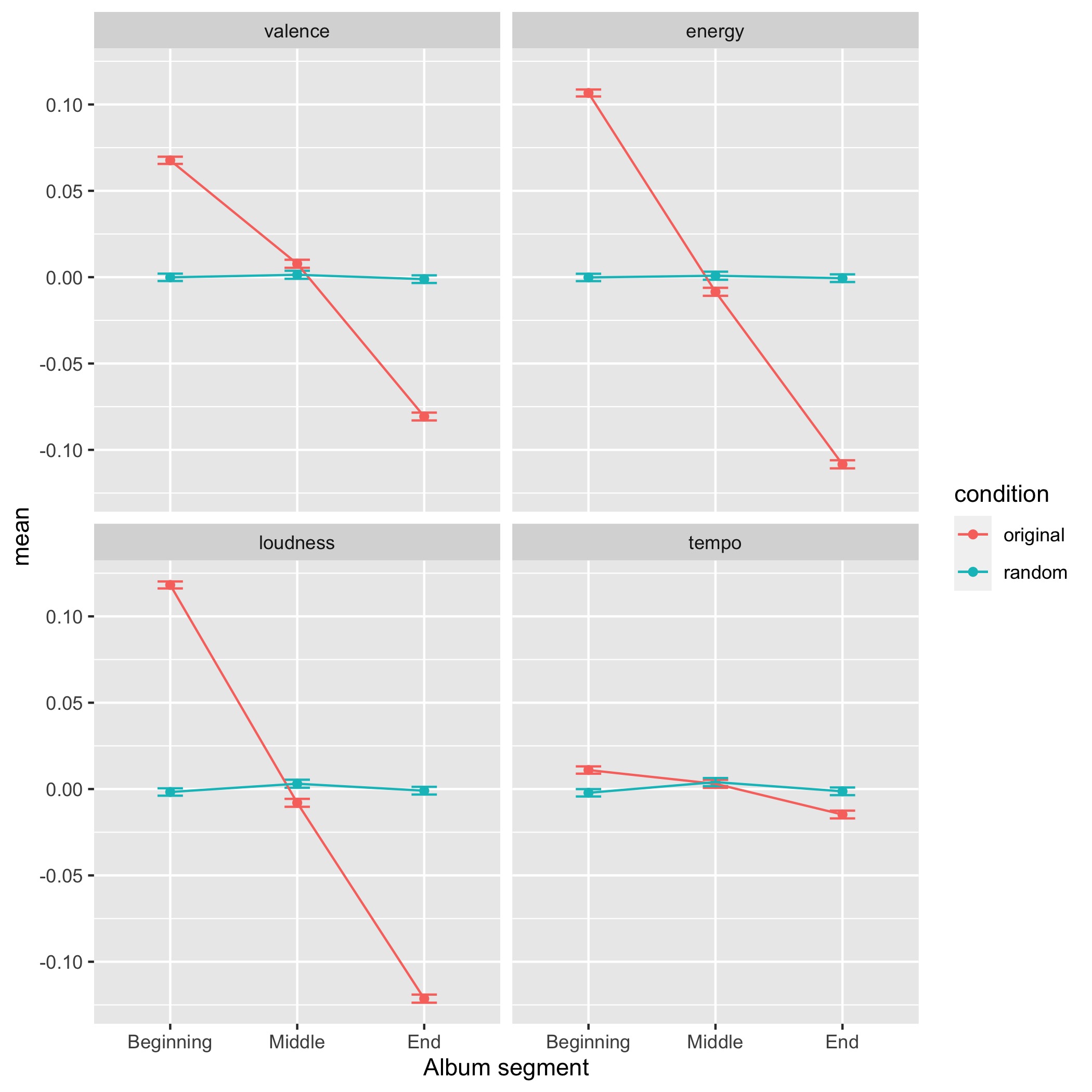}}
 \caption{Feature values throughout album segments. Error bars represent Standard Error of the Mean.}
 \label{fig:fig3}
\end{figure}

Our optimization was conducted with the method of Simulated Annealing (SA). We use the function $perm(a', q)$, which provides random permutations of $q$ tracks from $a'$. Note that the initial input is not $a$, as this would theoretically represent the best solution to our ordering problem. The output of $perm(a', q)$ generates the alternative permutation $a''$, and the probability of accepting $a''$ over $a'$ is a function of the improvement $\Delta\mathcal{L} = \mathcal{L}(a''|Tm)-\mathcal{L}(a'|Tm)$, as well as of the system's temperature $Temp$:

 \[ Pr(a''|a') = 
    \begin{cases}
      1, & \text{if}\ \Delta\mathcal{L} \geq 0 \\
     \\ exp(\frac{\Delta\mathcal{L}}{Temp}), & \text{otherwise}
    \end{cases}
  \]
 
We repeat this process with periodic decrements of both $Temp$ and $q$, which means that, as likelihood increases, there is a smaller probability of accepting $a''$ over $a'$ when $\Delta\mathcal{L} < 0$, as well as in fewer tracks being permuted.

\subsection{Evaluation}

To evaluate the performance of our system, we could compute the extent to which optimized albums are similar to the original ones. A standard accuracy test would count the frequency with which track number 1 of the original album appeared in position 1 of the optimized album. In this case, a perfect score would be obtained if the optimized album presented the exact same sequence of tracks as the original one. Notice that equation 2 cannot be used to evaluate our optimized albums, since this is already used as an objective function to find permutations of the tracks.

An alternative (and less rigorous) evaluation method would be to compute the number of correct pairs of tracks within each album. In this approach, we would compute an accurate response every time our optimization method indicated that track $k$ is more likely to be followed by track $k+1$. Notice that, with this approach, it does not matter if the optimized album starts with track number 5, as long as the next track is track number 6. The accuracy score would then be the frequency with which correct pairs occurred in our optimized albums. 

\begin{figure}
 \centerline{
 \includegraphics[width=0.7\columnwidth]{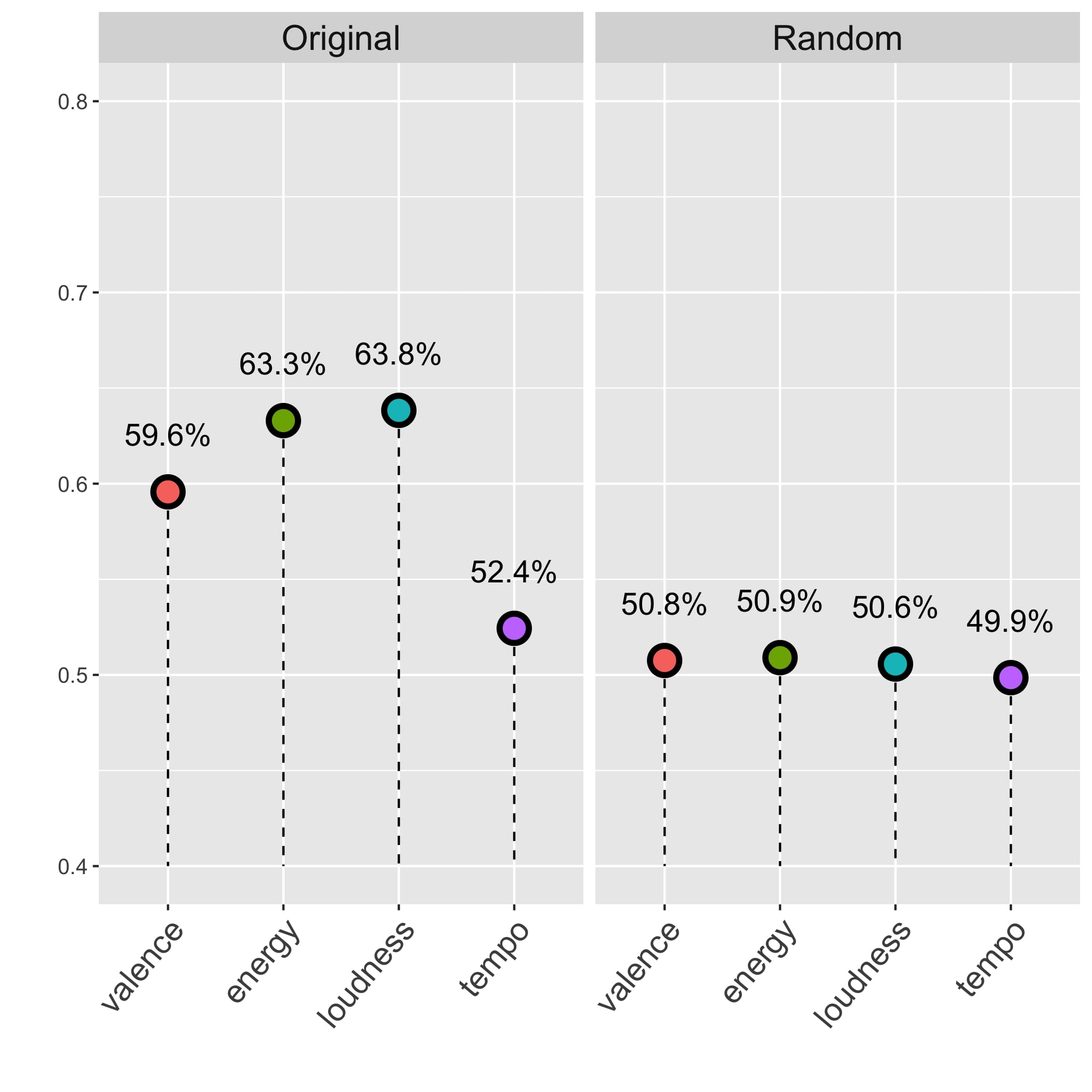}}
 \caption{Proportions of down ramps for original and randomized albums.}
 \label{fig:fig4}
\end{figure}

Given the large permutation spaces imposed by our problem, We chose to go with the less rigorous approach, and we tested the hypothesis that optimized albums would present more correct pairs of tracks than randomly permuted albums. By means of bootstrapping, we derived an empirical distribution of correct pairs for random permutations of all albums in our test dataset. We ran 10 thousand random permutations and computed the mean of immediate sequences within each run. This allowed us to derive sample means and its respective standard deviation, which approximates the standard error statistic. 

This empirical distribution allowed us to compute a one-sample t-test between 1) the number of sequences yielded by our optimized albums, and 2) the statistics derived from our bootstrap procedure. A good result would be obtained where the optimized albums contained a significantly higher number of sequences than the random permutations.

\subsection{Automated Track Sequencing Results}

\begin{figure}[h]
 \centerline{
 \includegraphics[width=0.9\columnwidth]{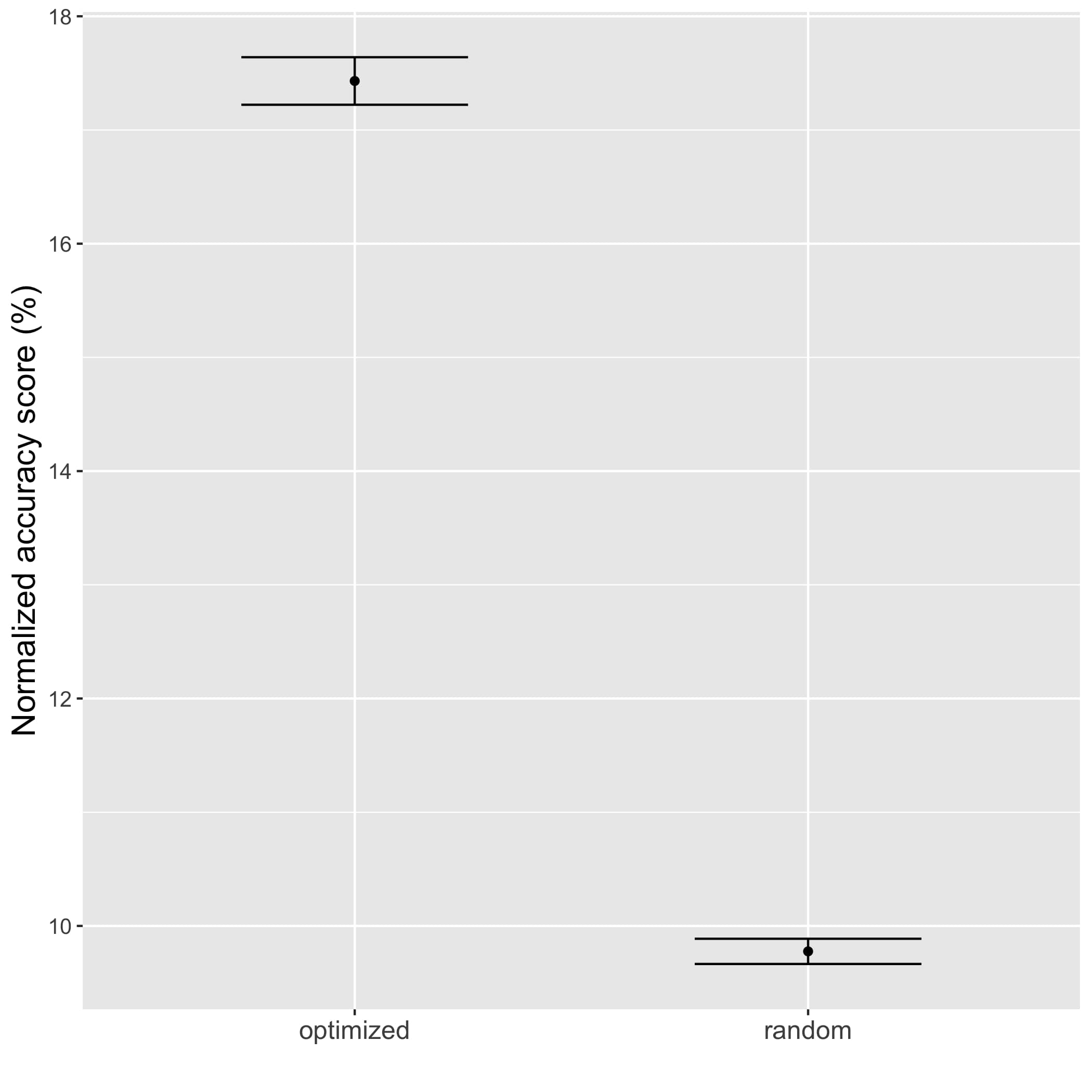}}
 \caption{Normalised accuracy score represents the number of correct sequences divided by the length of the album. Our system presents 17\% of correct sequences, whereas the random bootstrapped model presents an average of 9\%.}
 \label{fig:fig5}
\end{figure}

Our bootstrap t-test revealed a significant difference between random ($M = 0.88, SEM = 0.009$) and optimized ($M=1.61,\ SEM=0.02$) permutations [$t(10336)=35.69, CI=1.57-1.65,\ p<.001$] (\hyperref[fig:fig5]{Figure 5}). This indicates that our approach offers an above-chance level of finding the original next-track within an album.

\section{Discussion}

\subsection{Main findings}

Based on an overview of anecdotal beliefs \citep{nills2015art, ruoff2018how, ruth2019how, garvey2013art, hahn2018album, wensem2016how, ruoff2018how, sawyer2021how, friedman2021how}, we identified three basic principles of album-sequencing, namely 1) adjacent transitioning, 2) absolute positioning, and 3) overall trajectory. These principles were, to some extent, found in our dataset, which indicates that albums are constructed with some level of attention to sequential factors.

Regularities of \textbf{adjacent transitioning} were revealed by our transition matrices (Table 2), which suggested that changes of direction (i.e., up to down; down to up) are more probable than maintenance of direction (e.g., consecutive increases in loudness). This agrees with the results of \cite{nills2015art}, but is in slight opposition to the idea of maximum similarity expressed in the APG literature \citep{bittner2017automatic, flexer2008playlist, pohle2005generating, pohle2007reinventing, platt2001learning, kamehkhosh2017user, ikeda2017music, jannach2015beyond}. At least with regards to direction of feature values, change seems to be the norm.

Our transition matrices (Table 2) are indicative of album-sequencing regularities to the extent that 1) transition probabilities differ from chance levels; and 2) mean log-likelihoods (\hyperref[eqn:eq1]{eq. 2}) are higher for original albums than for randomized ones (\hyperref[fig:fig2]{Figure 2}). Regarding transition matrices, we note that probabilities are slightly above chance levels, but that they generally indicate a higher chance of changing directions between consecutive tracks (i.e., if a value goes up from track $k-1$ to track $k$, then it will most likely go down in track $k+1$).

One weakness of our adjacent transitioning approach is that it completely negates an estimation of transition magnitudes. Basically, our Markov model does not inform us about the extent to which consecutive tracks go up or down. Rather, it only informs us about the direction which consecutive tracks usually take. A closer look at transition magnitudes is arguably an interesting next step towards a deeper understanding of adjacent transitioning regularities.

On a global level, we showed that album sequences tend to prefer down ramps to up ramps (\hyperref[fig:fig4]{Figure 4}). This finding is consistent with the view that albums should follow an \textbf{overall trajectory}. Notice that the prevalence of down ramps is not inconsistent with our adjacent transitioning findings. As long as down transitions are higher in magnitude than up transitions, the album can have the same number of ups and downs, but still show a linear downtrend in a given feature domain.

This finding also relates to a set of studies conducted by \citep{huron1990crescendo, huron1991ramp, huron1992ramp, dean2008there, dean2010rise}, in which it was found that up and down ramps are not equally distributed within western compositions. For instance \cite{huron1991ramp} found that classical music presents more \textit{crescendi} than \textit{diminuendi}. \cite{dean2008there} and \cite{dean2010rise} presented contrasting results, indicating that down ramps are more frequent within a specific set of electroacoustic compositions. 

The loudness patterns that we found go against \cite{huron1990crescendo, huron1991ramp} and in favor of \cite{dean2008there, dean2010rise}. It should be noted, however, that our time-frames are much larger than the time-frames used in these previous studies, which looked for dynamic changes between measures of single compositions. Still, it is interesting to note that different increasing/decreasing trends might happen in different time-scales, also in different musical contexts. Whereas up ramps might be more frequent within a single classical composition \citep{huron1991ramp}, and still be used for particular cases within MAs, our sample showed a higher frequency of down ramps.

Concerning \textbf{absolute positioning}, our findings indicate that tracks with high levels of valence, energy and loudness are more likely to be positioned at the beginning of an album (\hyperref[fig:fig3]{Figure 3}). This suggests that musicians and album producers generally agree that the first segment of an album should have something different from the remaining segments \citep{hahn2018album, wensem2016how, ruoff2018how}. We note, however, that there are nothing but vague suggestions regarding how albums should begin. It is usually said, for instance, that the best songs should be placed in the beginning, but we still do not know what it means for a song to be the best one.

Conceptually, slow and quiet songs could be viewed as the ``best" ones, and therefore be positioned at the beginning of the albums. Still, this is not what our data shows. If we consider that our albums tended to start with high levels of valence, energy and loudness, we could hypothesize that artists see these features as being indicative of quality, or maybe as indicative of the subjective value that listeners will assign to it. Still, we leave an open question: why is it better to start an album with high levels of valence, energy and loudness?

Finally, we showed that album-tracks can be sorted algorithmically. If we optimize parameters of adjacent transitioning and overall-trajectory, we end up with albums that present more direct increases (e.g., track $k+1$ comes after track $k$), than if we randomly permutate these albums. Future venues of research might incorporate deep-learning, alternative optimization techniques, as well as more robust evaluation procedures, such as ablation studies and perceptual validation.

\subsection{Perceptual implications - global or local?}

With our computational approach, we have no direct means for drawing conclusions about how album-sequences are perceived. We do suggest, however, that musicians and album producers agree, to some extent, about the order in which songs should be presented to the listener. If we assume that albums are assembled by intelligent agents who are interested in optimizing some dimension of the musical experience, we could still hypothesize that track ordering might have some degree of perceptual relevance.

This assumption is strengthened by some studies suggesting that the perception of sound intensity (i.e., loudness) is heavily shaped by time-related factors. \cite{patterson1974musical}, \cite{geringer1995continuous}, \cite{geringer2003gradual} and \cite{olsen2014intensity} show, for instance, that sequential manipulations of sound intensity, such as sudden and abrupt increases in Sound Pressure Level, can have a significant effect on listener's attention, chills, and levels of arousal. The phenomenon known as loudness adaptation \citep{kimura2004effects, dange1993loudness}, on the other hand, indicates that the perception of intensity decreases with time, even if the stimulus is held at a constant SPL. This shows that not only the immediate physical properties of the stimulus is capable of influencing how a sound is perceived, but also the context in which the stimulus is inserted.

In the context of our study, the effect of loudness adaptation can help us explain why consecutive tracks were constantly going up and down. If true that the perception of sound-intensity can decrease as much as 50\% after 3 minutes of exposure \citep{miskiewicz1993loudness}, it might also be the case that musicians try to counterbalance this effect by alternating between increases and decreases of loudness throughout the album (\hyperref[fig:fig3]{Figure 3}). 

Concerning the effect of overall-trajectory (\hyperref[fig:fig4]{Fig. 4}) we hypothesize that down ramps are a mere by-product of the absolute positioning effect (\hyperref[fig:fig3]{Fig. 3}). It might be the case that, by opening the albums with high levels of valence, energy and loudness, the only tracks that remain for the last segment are those with lower values of these features. In this case, gradual decreases throughout the album might not reflect any aesthetic or perceptual purpose, but just the fact that first positions of an album are reserved for songs which are more likely to impact the listener. 

\begin{figure}[h]
 \centerline{
 \includegraphics[width=0.9\columnwidth]{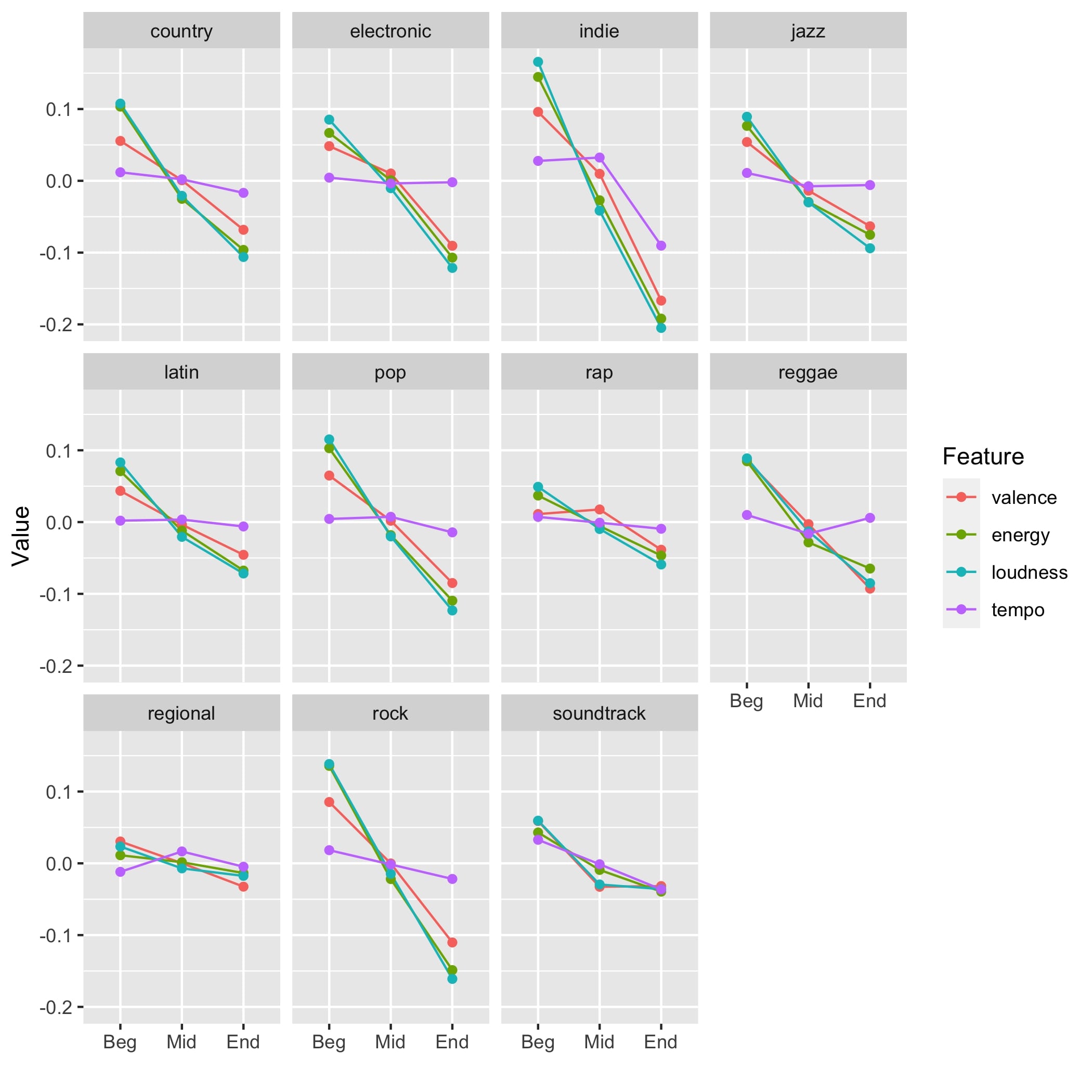}}
 \caption{Normalised feature values across different segments of the albums, separated by genre.}
 \label{fig:fig6}
\end{figure}

We also acknowledge the possibility that our findings have no perceptual basis, and that listeners are generally oblivious to sequential factors when it comes to track-ordering. This possibility is strengthened by the data which supports a concatenationist view of music perception \citep{konevcni1984elusive, tillmann2004relative, tillmann2006cognitive, eitan2008growing, gotlieb1985effects, levinson2006concatenationism, levinson2018music}. However, as previously argued, the hypothesis of global perception is usually falsified by studies focusing on classical repertoire, and on highly abstract and idiomatic features, such as tonal closure and motivic development.

Finally, we refrained from including genre as an independent variable in our analyses, since we had no working hypothesis to test. Throughout our review of anecdotal beliefs, there was no mention of different approaches for genre. Also, the psychological effect of loudness adaptation---or any other psychoacoustic phenomenon that is relevant to how we perceive songs in a sequence---should be consistent between different genres. In fact, \hyperref[fig:fig6]{Figure 6} shows that feature distributions are consistent throughout album segments of different genres. Still, these questions may be investigated in future studies.

\section*{Funding}

This work was supported by the Finnish National Agency for Education and by the Center of Excellence in Music, Mind, Body and Brain through the Research Council of Finland.

\bibliography{interactapasample}


\end{document}